\input{aipcheck}

\newcommand{\be}{\begin{equation}}
\newcommand{\ee}{\end{equation}}
\newcommand{\Deln}{\ensuremath{\Delta N_\nu\;}}
\newcommand{\epm}{\ensuremath{e^{\pm}\;}}

\def\eg{{\it e.g.},~}
\def\etal{{\it et al.}~}
\def\4he{$^4$He}
\def\3he{$^3$He}
\def\7li{$^7$Li}
\def\Yp{Y$_{\rm P}$~}
\def\yd{$y_{\rm D}$~}
\def\hii{H\thinspace{$\scriptstyle{\rm II}$}~}
\def\hii{H\thinspace{$\scriptstyle{\rm II}$}~}
\def\hi{H\thinspace{$\scriptstyle{\rm I}$}~}  
\def\di{D\thinspace{$\scriptstyle{\rm I}$}~}  
\def\Nnu{N$_{\nu}$~}

\newcommand\la{\lower0.6ex\vbox{\hbox{\ensuremath{\buildrel{\textstyle<}\over{\sim}\ }}}}
\newcommand\ga{\lower0.6ex\vbox{\hbox{\ensuremath{\buildrel{\textstyle>}\over{\sim}\ }}}}
\newcommand{\obh}{\ensuremath{\Omega_{\rm B} h^2\;}}

\documentclass[
    ,final            
  ]
  {aipproc}

\layoutstyle{6x9}

\begin{document}

\title{BBN And The CBR Probe The Early Universe}

\author{Gary Steigman}{
  address={Departments of Physics and Astronomy, The Ohio State University,
191 West Woodruff Avenue, Columbus, OH 43210, USA}
}

\begin{abstract}
Big Bang Nucleosynthesis (BBN) and the Cosmic Background Radiation (CBR)
provide complementary probes of the early evolution of the Universe and 
of its particle content.  Neutrinos play important roles in both cases,
influencing the primordial abundances of the nuclides produced by BBN
during the first 20 minutes as well as the spectrum of temperature
fluctuations imprinted on the CBR when the Universe is some 400 thousand
years old.  The physical effects relevant at these widely separated epochs 
are reviewed and the theoretical predictions are compared with observational 
data to explore the consistency of the standard models of cosmology and 
particle physics and to constrain beyond-the-standard-model physics and
cosmology.
\end{abstract}

\maketitle

\section{Introduction}

The Universe is expanding and is filled with radiation.  All wavelengths, 
of photons as well as the deBroglie wavelengths of freely expanding 
massive particles, are stretched along with this expansion.  As a 
result, during its earlier evolution the Universe was hot and dense.  
The combination of high temperature and density ensures that collision 
rates are very high during early epochs, guaranteeing that all particles, 
with the possible exception of those with only gravitational strength 
interactions, were in equilibrium at sufficiently early times.  As the 
Universe expands and cools, interaction rates decline and, depending 
on the strength of their interactions, different particles depart from 
equilibrium at different epochs.  For the standard, ``active" neutrinos 
($\nu_{e}$, $\nu_{\mu}$, $\nu_{\tau}$) departure from equilibrium occurs 
when the Universe is only a few tenths of a second old and the temperature 
of the CBR photons, \epm pairs, and the neutrinos is a few MeV.  It should 
be emphasized that departure from equilibrium is not sharp and collisions 
continue to occur.  For $T~\la 2-3$~MeV, the neutrino interaction rates 
become slower than the universal expansion rate (as measured by the Hubble 
parameter $H$), effectively decoupling the neutrinos from the CBR photons 
and \epm pairs present at that time.  However, electron neutrinos (and 
antineutrinos) continue to interact with the baryons (nucleons) 
via the charged-current, weak interactions until the Universe is a few 
seconds old and the temperature has dropped below an MeV.  Once again, this 
decoupling is not abrupt (the neutrinos do {\it not} ``freeze-out").  Two body 
reactions among neutrons, protons, \epm and $\nu_{e}$($\bar{\nu}_{e}$) 
continue to influence the ratio of neutrons to protons, albeit not rapidly 
enough to allow the n/p ratio to track its equilibrium value of n/p = 
exp$(-\Delta m/T)$, where $\Delta m = m_{\rm n} - m_{\rm p} = 1.29$~MeV.  
As a result, the n/p ratio decreases from $\sim 1/6$ at ``freeze-out" to 
$\sim 1/7$ when BBN begins at $\sim 200$~sec ($T \approx 80$~keV).  Since 
the neutrinos are extremely relativistic during these epochs, they can 
influence BBN in several ways.  The universal expansion rate in the standard 
cosmology is determined through the Friedman equation by the total energy 
density which is dominated during these early epochs by massless particles 
along with those massive particles which are extremely relativistic at 
these epochs: CBR photons, \epm pairs, neutrinos.  The early Universe is 
``radiation" dominated and neutrinos constitute a significant component 
of the ``radiation".  In addition, through their charged-current weak 
interactions the electron-type neutrinos help to control the neutron-to-proton 
ratio, effectively limiting the primordial abundance of \4he.

Although \epm pairs annihilated during the first few seconds, the surviving
electrons, equal in number to the protons to ensure charge neutrality, are
coupled to the CBR photons via Compton scattering.  Only after the electrons 
and nuclides (mainly protons and alphas) combine to form neutral atoms 
(``recombination") are the CBR photons released from the grasp of the 
electrons to propagate freely.  This occurs when the Universe is some 400 
thousand years old and the relic photons, redshifted to the currently 
observed black body radiation at $T = 2.725$K, provide us a snapshot of 
the universe at this early epoch.  At this relatively late stage (compared 
to BBN) in the early evolution of the Universe, the key role of the freely 
propagating, relativistic neutrinos is their contribution to the total 
radiation density, which determines the universal expansion rate (\eg the 
time -- temperature relation).  It should be noted that if the neutrino 
masses are sufficiently large the neutrinos will have become nonrelativistic 
and their free-streaming has the potential to damp density fluctuations 
in the baryon fluid.  This important topic is not addressed here.
   
The primordial abundances of the relic nuclei produced during BBN depend 
on the baryon (nucleon) density and on the early-Universe expansion rate.  
The amplitudes and angular distribution of the CBR temperature fluctuations
depend on these same parameters (as well as several others).  The universal 
abundance of baryons may be quantified by comparing the number of baryons 
(nucleons) to the number of CBR photons,
\be
\eta_{10} \equiv 10^{10}(n_{\rm B}/n_{\gamma}).
\ee
As the Universe expands the densities of baryons and photons both decrease 
but the numbers of baryons and of CBR photons in a comoving volume are 
unchanged (post-\epm annihilation) so that $\eta_{10}$ measured at present, 
at recombination, and at BBN should all be the same.  This is one of the 
key cosmological tests.  Since the baryon mass density ($\rho_{\rm B} \equiv 
\Omega_{\rm B} \rho_{c}$, where $\rho_{c} = 3H_{0}^{2}/8\pi G$ is the present 
critical mass density) plays a direct role in the growth of perturbations, 
it is convenient to quantify the baryon abundance using a combination of 
$\Omega_{\rm B}$ and $h$, the present value of the Hubble parameter ($H_{0}$) 
measured in units of 100 kms$^{-1}$Mpc$^{-1}$, 
\be
\eta_{10} = 274~\omega_{\rm B} \equiv 274~\Omega_{\rm B}h^{2}.
\ee
The Hubble parameter, $H = H(t)$, measures the expansion rate of the Universe.
Deviations from the standard model ($H\rightarrow H'$) may be parameterized 
by an expansion rate parameter $S\equiv H'/H$.  Since $H$ is determined in the 
standard model by the energy density in relativistic particles, deviations 
from the standard cosmology ($S \neq 1$) may also be quantified by the 
``equivalent number of neutrinos" \Deln $\equiv N_{\nu} - 3$.  Prior to \epm 
annihilation, these two parameters are related by
\be
S = (1 + 7\Delta N_{\nu}/43)^{1/2}.
\ee
\Deln is an equivalent and convenient way to quantify {\it any} deviation from 
the standard model expansion rate ($S$); it is not necessarily related to 
extra (or fewer!) neutrinos.

The question considered here is, ``Are the predictions and observations 
of the baryon density and the expansion rate of the Universe at 20 minutes 
(BBN) and 400 thousand years (CBR) in agreement with each other and with 
the standard models of cosmology and particle physics?".  If yes, what 
constraints are there on models of beyond-the-standard-model physics and/or 
cosmology?   The current status of this quest is summarized here.  For more 
detail and further references, see my recent review article~\cite{bbnrev}.

\section{The Universe At 20 Minutes: BBN}
Nuclear reactions among nucleons occur rapidly during the early evolution 
of the Universe but they fail to produce significant abundances of complex 
nuclides because of their competition with photo-destruction reactions 
involving the enormously more abundant CBR photons.  When the Universe 
cools sufficiently ($T~\la 80$~keV, $t~\ga 3$~minutes), reducing the number 
of photons capable of photo-destruction, BBN begins in earnest and the 
available neutrons are consumed very quickly to build \4he.  All further 
nucleosynthesis among electrically charged nuclides (H, D, T, \3he, \4he) 
involves reactions which become Coulomb-supressed as the Universe expands 
and cools.  As a result BBN terminates when $T~\la 30$~keV ($t~\ga 25$~min).  
In the first $\sim 20$~minutes of its evolution the cosmic nuclear reactor 
produces (in astrophysically interesting abundances) deuterium, helium-3 
(any tritium decays to \3he), helium-4 and, because of the gaps at mass-5 
and mass-8, only a tiny amount of lithium-7 (produced mainly as beryllium-7 
which, later in the evolution captures an electron and decays to \7li).

\begin{figure}[h]
\includegraphics[width=20pc]{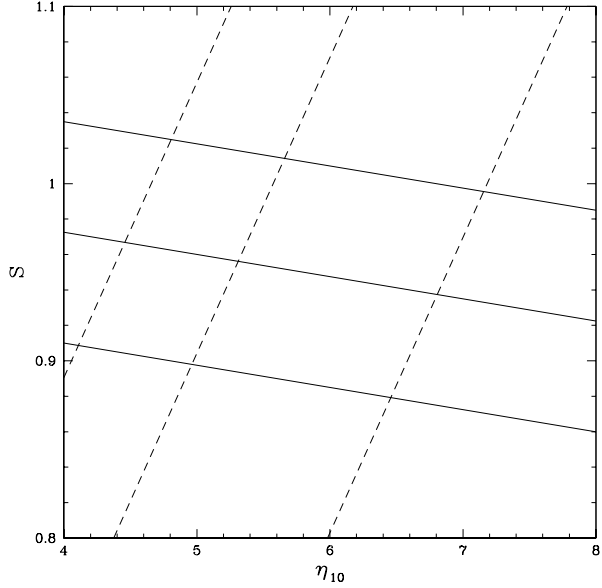}
\caption{\label{fig:svseta}Isoabundance curves for Deuterium (dashed lines) 
and Helium-4 (solid lines) in the expansion rate factor ($S$) -- baryon 
abundance ($\eta_{10}$) plane.  The \4he curves, from bottom to top, are 
for \Yp = 0.23, 0.24, 0.25.  The D curves, from left to right, are for
\yd = 4.0, 3.0, 2.0.}
\end{figure}

The BBN-predicted abundances of D, \3he, and \7li are determined by the 
competition between production and destruction rates which depend on the 
overall density of baryons. As a result these nuclides are all potential 
baryometers.  Among them, D is the baryometer of choice since its post-BBN 
evolution is simple (when gas is incorporated into stars D is only destroyed, 
burned to \3he and beyond) and the BBN-predicted primordial D abundance 
is a relatively sensitive function of the baryon density (D/H $\propto 
\eta_{10}^{-1.6}$).  In contrast, the primordial abundance of \4he is 
relatively insensitive to the baryon density, controlled mainly by the 
abundance of neutrons when BBN begins.  The \4he relic abundance is 
usually expressed as a ``mass fraction" \Yp $\equiv 4y/(1+4y)$, where $y 
\equiv n_{\rm He}/n_{\rm H}$ (since this {\it assumes} $m_{\rm He}/m_{\rm H} 
= 4$, \Yp is {\it not} the true helium {\it mass} fraction).  Since the 
expansion rate ($S$), in combination with the rate of the charged-current 
weak interactions, plays an important role in regulating the pre-BBN 
neutron to proton ratio, \Yp is sensitive to $S$.  As shown by the D and 
\4he isoabundance curves in Figure~\ref{fig:svseta}, deuterium and helium-4 
provide complementary probes of the universal baryon density and expansion 
rate parameters.

For restricted but interestingly large ranges of $\eta_{10}$($\omega_{\rm
B}$), $\Delta N_{\nu}$($S$), Kneller and Steigman~\cite{ks} found simple but 
accurate analytic fits to the BBN-predicted abundances of the light nuclides 
as functions of $S$ and $\eta_{10}$.  For D ($y_{\rm D} \equiv 10^{5}$(D/H)) 
and \4he (Y$_{\rm P}$), these are
\be
y_{\rm D} \equiv 46.5(1 \pm 0.03)\eta_{\rm D}^{-1.6}~; ~~{\rm Y}_{\rm P} \equiv 
(0.2384 \pm 0.0006) + \eta_{\rm He}/625,
\ee
where
\be
\eta_{\rm D} \equiv \eta_{10} - 6(S-1)~; ~~\eta_{\rm He} \equiv \eta_{10} + 100(S-1).
\ee

\subsection{Observed Relic Abundances}
\begin{figure}[h]
\includegraphics[width=20pc]{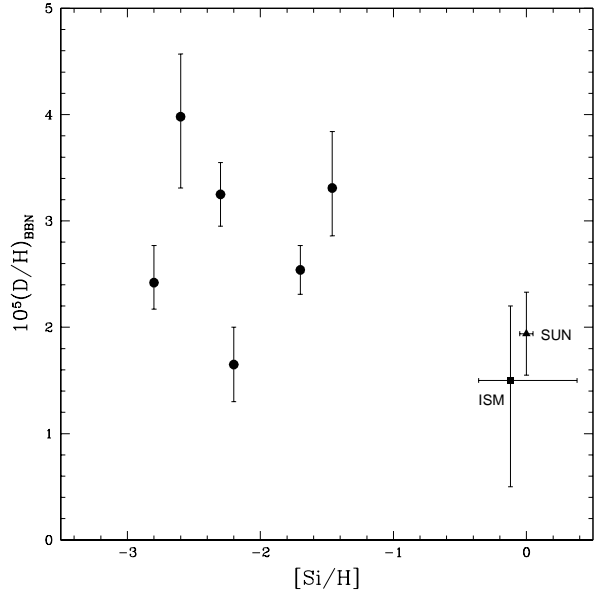}
\caption{\label{fig:dvssi06}Observationally inferred deuterium 
abundances versus metallicity for six high redshift, low 
metallicity QSOALS (filled circles).  Also shown are the 
abundances derived for the pre-solar nebula (Sun) and for 
the local interstellar medium (ISM).}
\end{figure}

Observations of deuterium in the solar system and the interstellar medium 
(ISM) of the Galaxy provide interesting {\it lower} bounds to its primordial 
abundance but it is the handful of observations of D in high redshift, low 
metallicity, QSO absorption line systems (QSOALS) which are of most value 
in providing estimates of the primordial D abundance.  The identical 
absorption spectra of \di and \hi (modulo the velocity/wavelength shift 
resulting from the heavier reduced mass of the deuterium atom) is a 
liability, limiting the number of useful targets in the vast Lyman-alpha 
forest of QSO absorption spectra (see, \eg Kirkman {\it et al.}~\cite{kirk} 
for further discussion).  Until recently there were only five QSOALS with 
deuterium detections leading to reasonably reliable abundance determinations 
\cite{kirk} (and references therein); these, along with a very recent sixth 
determination by O'Meara \etal \cite{omeara}, are shown in Figure~\ref{fig:dvssi06}.  
Also shown there for comparison are the solar system (pre-solar nebula) 
and ISM D abundances.  Clearly there is excessive dispersion among the 
low metallicity D abundance determinations, suggesting that systematic 
errors, whose magnitudes are hard to estimate, may have contaminated the 
determinations of at least some of the \di and/or \hi column densities.  
This dispersion serves to mask the anticipated primordial deuterium plateau.  
The best that can be done with the present data is to identify the relic 
deuterium abundance with the weighted mean of the high-$z$, low-$Z$ D/H 
ratios: $y_{\rm D} \equiv 2.68^{+0.27}_{-0.25}$, corresponding to 
$\eta_{\rm D} = 5.95^{+0.36}_{-0.39}$ (see eq.~4).
\begin{figure}[h]
\includegraphics[width=20pc]{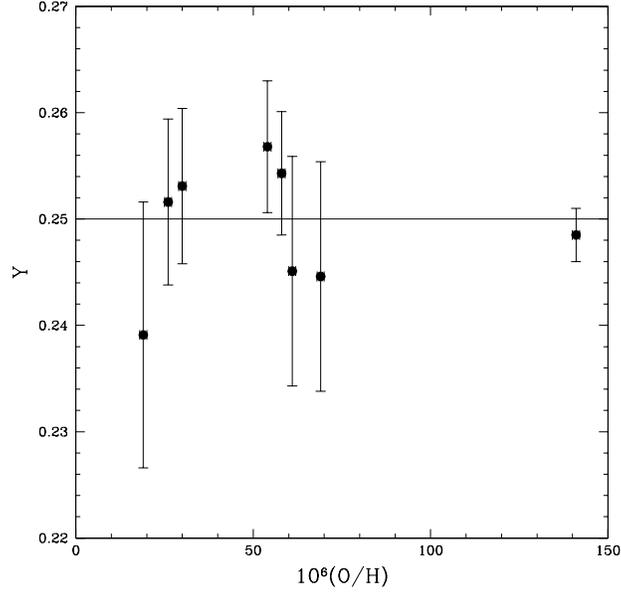}
\caption{\label{fig:hevsoos}The OS-derived \4he versus oxygen
abundances for 7 low metallicity \hii regions from IT 
and one higher metallicity \hii region from Peimbert \etal 
(filled circles).  The horizontal line shows the weighted 
mean of the 8 helium abundances.}
\end{figure}

The post-BBN evolution of \4he is also quite simple (monotonic).  
As gas cycles through generations of stars, hydrogen is burned to 
helium-4 (and beyond), increasing the \4he abundance above its 
primordial value.  As a result, the present \4he mass fraction, 
Y$_{0}$, has received a significant contribution from post-BBN, 
stellar nucleosynthesis, and Y$_{0} >$~Y$_{\rm P}$.  However, since 
the ``metals" such as oxygen are produced by short-lived, massive 
stars and \4he is synthesized (to a greater or lesser extent) by 
stars of all masses, at very low metallicity the increase in Y 
should lag that in, \eg O/H, so that as O/H $\rightarrow 0$, Y 
$\rightarrow$ Y$_{\rm P}$.  Therefore, although \4he is observed 
in the SuFieldsn and in Galactic \hii regions where the metallicity is 
relatively high, the crucial data for inferring its primordial 
abundance are provided by observations of helium and hydrogen 
emission (recombination) lines from low-metallicity, extragalactic 
\hii regions.  The present inventory of such regions studied for 
their helium content exceeds 80 (see Izotov \& Thuan (IT)~\cite{it}).  
Since for such a large data set even modest observational errors 
for the individual \hii regions can lead to an inferred primordial 
abundance whose {\it formal} statistical uncertainty is very small, 
special care must be taken to include hitherto ignored systematic 
corrections and/or errors.  It is the general consensus that the 
present uncertainty in \Yp is dominated by the latter, rather than 
by the former errors.  However, attempts to include estimates of 
systematic errors have often been unsystematic or absent.  To 
account for some of these uncertainties, Olive, Steigman, and 
Walker~\cite{osw} adopted the Fields and Olive~\cite{fo} analysis to 
estimate a $\sim 95\%$ confidence range of $0.228 \leq {\rm Y}_{\rm P} 
\leq 0.248$ (if it is assumed that Y$_{\rm P} = 0.238 \pm 0.005$ 
($\sim 1\sigma$), this corresponds to $\eta_{\rm He} = -0.25 \pm 3.15$; 
see eq.~4).   The most systematic analysis of the IT data is that by 
Olive \& Skillman 2004 (OS)~\cite{os}.  Using criteria outlined in their 
earlier paper~\cite{os}, OS examined the IT data set, concluding that 
they could apply their analysis to only 7 of the 82 IT \hii regions.  
This tiny subset of the data, when combined with its limited range 
in oxygen abundance, severely limits the statistical significance 
of the OS conclusions.  In Figure \ref{fig:hevsoos} are shown the 
OS-inferred helium abundances from the IT data set and from one, 
higher metallicity \hii region observed by Peimbert \etal \cite{peim}.  
These eight \hii regions alone provide no evidence in support of a 
correlation of the helium abundance with the oxygen abundance.  The 
weighted mean is $<$Y$> = 0.250\pm0.002$, leading to a robust $\sim 
2\sigma$ upper bound on the primordial helium abundance of \Yp $\leq 
0.254$ (corresponding to $\eta_{\rm He} \leq 9.75$).

\subsection{Comparison Between Predicted And Observed Relic Abundances}

\begin{figure}
  \includegraphics[width=20pc]{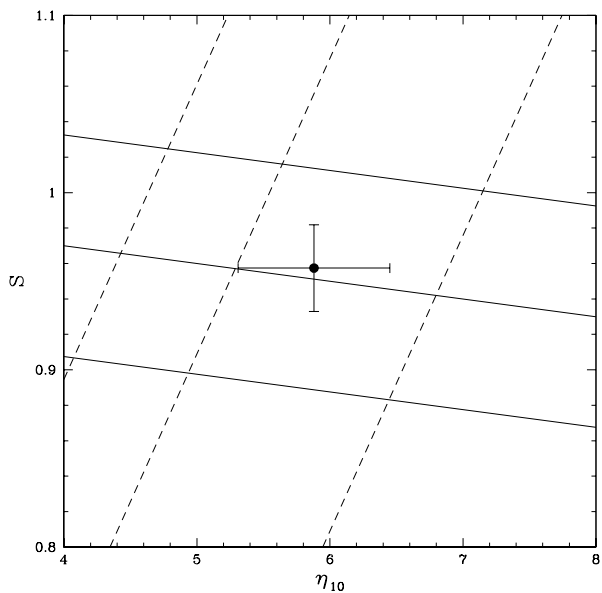}
  \caption{\label{fig:svsetadhe}Isoabundance curves for Deuterium (dashed 
lines) and Helium-4 (solid lines) in the expansion rate factor -- baryon 
abundance plane (see Fig.~\ref{fig:svseta}).  The filled circle and 
error bars correspond to the adopted D and \4he relic abundances.}
\end{figure}
The relic abundances adopted here correspond to $\eta_{\rm D} = 
5.95^{+0.36}_{-0.39}$ and $\eta_{\rm He} = -0.25 \pm 3.15$, corresponding 
to $S = 0.942\pm0.030$ (N$_{\nu} = 2.30^{+0.35}_{-0.34}$) and $\eta_{10} 
= 5.60^{+0.38}_{-0.41}$ (\obh $= 0.0204^{+0.0014}_{-0.0015}$).  This is 
consistent with the standard model ($S = 1$, \Nnu = 3) at $\sim 2\sigma$ 
(see Figure~\ref{fig:svsetadhe}).  The confrontation of the BBN predictions 
with the relic abundance observations of D and \4he reveals internal 
consistency (at $\la 2\sigma$) of the standard models of particle physics 
(N$_{\nu} = 3$) and cosmology ($S = 1$) and it fixes the baryon abundance 
to an accuracy of $\sim 7\%$ during the first few minutes of the evolution 
of the Universe.  At the same time this comparison sets constraints on 
possible deviations from these standard models (\eg N$_{\nu} \neq 4$).  How 
do these BBN results compare with what the CBR reveals about the Universe 
some 400 thousand years later?

\section{Confrontation With The CBR}

The angular spectrum of CBR temperature fluctuations depends on 
several key cosmological parameters, including the baryon density 
and the relativistic energy density (for further discussion and 
references, see Hu and Dodelson 2002~\cite{hu} and Barger \etal 2003 
\cite{barger}), thereby providing a probe of $\eta_{10}$ and N$_{\nu}$ 
some 400 kyr after BBN.  With \Nnu allowed to depart from the standard 
model value, Barger \etal \cite{barger} found the first year WMAP data
\cite{sperg03} is best fit by \obh = 0.0230 and \Nnu = 2.75, in excellent 
agreement with the purely BBN results above.  In fact, the CBR is a 
much better baryometer than it is a chronometer, so that while the 
$2\sigma$ range for the baryon density is limited to $0.0204 \leq 
\Omega_{\rm B}h^{2} \leq 0.0265$, the corresponding $2\sigma$ range 
for \Nnu was found to be $0.9 \leq {\rm N}_{\nu} \leq 8.3$~\cite{barger}. 

Quite recently the WMAP team released (and analyzed) their 3-year 
data.  For \Nnu = 3, Spergel \etal 2006~\cite{nnu} find $\Omega_{\rm B}h^{2}
= 0.0223^{+0.0007}_{-0.0009}$.  With \Nnu free to vary, V. Simha and 
the current author, in preliminary work in progress, find a similar 
result for the baryon density (not unexpected since in fitting to the 
CBR data the contributions from the baryon density and the relativistic 
energy density (or, $S$) are largely uncorrelated), $\Omega_{\rm B}h^{2} 
= 0.0222\pm0.0007$, along with a $2\sigma$ range for \Nnu ($0.7~\la 
{\rm N}_{\nu}~\la 6.0$) which is somewhat smaller than the previous 
WMAP-based result~\cite{barger}.  However, this result is preliminary and 
should be regarded with a very large grain of salt (caveat emptor!).
 
\section{Summary}

Comparison between the BBN predictions and relic abundance observations 
of deuterium and helium-4 reveals consistency with the standard models 
of particle physics and cosmology and constrains the value of the baryon 
abundance during the first few minutes of the evolution of the Universe.  
This comparison also enables quantitative constraints on possible deviations 
from these standard models, particularly in the neutrino sector.  Some 
400 thousand years later, when the CBR photons are set free, the angular 
spectrum of temperature fluctuations encodes information about several 
key cosmological parameters, including \Nnu and $\Omega_{\rm B}h^{2}$.  
The present data reveal consistency (at $\sim 2\sigma$) between the 
values of \obh and \Nnu inferred from the first few minutes of the 
evolution of the Universe and from a snapshot of the Universe some 
400 kyr later.  While there is still room for surprises, at present 
the standard models appear robust.

\begin{theacknowledgments}
The author's research is supported at The Ohio State University by a 
grant (DE-FG02-91ER40690) from the US Department of Energy.
\end{theacknowledgments}



\bibliographystyle{aipproc}   

\bibliography{sample}

\IfFileExists{\jobname.bbl}{}
 {\typeout{}
  \typeout{******************************************}
  \typeout{** Please run "bibtex \jobname" to optain}
  \typeout{** the bibliography and then re-run LaTeX}
  \typeout{** twice to fix the references!}
  \typeout{******************************************}
  \typeout{}
 }

\end{document}